# On-chip ultrafast plasmonic graphene hot electron bolometric photodetector

*Jacek Gosciniak[1], and Jacob B. Khurgin[2]*

[1]New York University Abu Dhabi, Saadiyat Island, P.O. Box 129188, Abu Dhabi, UAE.
[2]John Hopkins University, Baltimore, MD 21218, USA

**Abstract**
We investigate waveguide-integrated plasmonic graphene photodetector operating based on the hot carrier photo-bolometric effect, which is characterized simultaneously by high responsivity on the scale of hundreds of A/W and high speed on the scale of 100's of GHz. We develop a theory of bolometric effect originating from the band nonparabolicity of graphene and estimate responsivity due to bolometric effect is shown to significantly surpass the responsivity of co-existing photo-conductive effect thus convincingly demonstrating the dominance of bolometric effect. Based on the theory we propose a novel detector configuration based on hybrid waveguide that allows for efficient absorption in the graphene over short distance and subsequently a large change of conductivity. The results demonstrate the potential of graphene for high-speed communication systems.

**Introduction**
On-chip high-speed photodetectors are crucial components in optical communications that convert the absorbed photons into an electrical signal [1, 2]. Being the last components in the optical links, the detectors must operate with low power if the costly amplifiers are to be avoided. Other requirements, such as they operating beyond 100 GHz and be characterized by low current and high responsivity. Most traditional optoelectronic devices suffer, however, from either a high power consumption, low operation speed, low responsivity or large footprint [3]. Thus, finding a better design platform is under a deep interest for industry and academia. In the last few decades it has become obvious that further progress can be made either through an implementation of new materials [4-6], integration with plasmonics [7-10] or both. Plasmonics can squeeze light much below a diffraction limit, which reduces the device footprint [8, 9]. Furthermore, the small device volume means a higher density of integration, and simultaneously, lower power consumption, easier heat dissipation, and faster operation speed [8, 11-13]. During the last few years some waveguide-integrated plasmonic photodetectors on silicon [12-18] and germanium [19, 20] have been proposed and fabricated. However, the search for better material platform continues. Graphene is a very promising material for signal modulation and photodetection owing to its extraordinary transport properties [21-23]. Being only one atom thick it absorbs 2.3 % of incident light in a very wide energy spectrum [5, 22, 24]. It has ultrafast carrier dynamics, tunable optical properties and high carrier mobility enabling ultrafast conversion of photons or plasmons to electrical current or voltage [25-27]. Vice versa, it can convert a received electrical signal onto an optical carrier in this way working as the optical modulator [5, 28-31]. Moreover, graphene is CMOS-compatible allowing integration on wafer-scale [3-5]. Graphene photodetectors can operate based on photovoltaic [11, 32], photo-thermoelectric [13, 33], photo-gaining [34, 35] or photo-bolometric [12, 32, 36-40] effects. The choice of effect depends on a photodetector's configuration and specific applications [37, 41]. Bolometers have emerged as the technology of choice, because they do not need cooling. Actually, they utilize the temperature-dependent material properties for photodetection - the incident light raises the local electronic temperature of the material, which reduces the resistance of the device and produces a change in the current [41-45]. Graphene is well-suited for this purpose as it has a small electron heat capacity and weak electron-lattice (e-l) coupling leading to a large light-induced change in electron temperature [13, 16, 25, 26]. The low density of states and small volume for a given area result in low heat capacity. Simultaneously, the cooling of electrons by acoustic phonons is inefficient, and cooling by optical phonons requires high lattice temperatures exceeding 2300 K, equivalent to phonon energy of 200 meV [46, 47]. As a result, electron temperature $T_e$ can rise to 1000's of Kelvins and thus engender strong bolometric response [13, 33].

Up to date, the state-of-the art graphene waveguide-integrated PB photodetectors operating at telecom wavelength and room temperature conditions show very promising performances with a responsivity exceeding 0.35 - 0.5 A/W [12, 32] and bandwidth above 100 GHz. However, there is still a lot of matter for improvement such as minimizing the absorption losses by metal, improvement of coupling efficiency, efficient heat transfer to the electrons and etc. Therefore, finding a better arrangement for the realization of graphene-based bolometers operating at room temperatures and telecom wavelengths is highly desired.

Developing efficient bolometric detector requires understanding full physical picture of how exactly heating of electrons affect conductivity. Usually it is done by assuming that electrons quickly thermalize with a given electron temperature, but in fact it has not been shown that the electron distribution can indeed be described by perfect Fermi Dirac distribution. Furthermore, specific of graphene is such that bolometric effect of increased resistance is always accompanied by the photoconductive effect causes by interband absorption that reduces resistance. In this work we develop a simple theory of bolometric effect in graphene that does not require establishment of a perfect equilibrium between the electrons and leads to a simple expression for responsivity that depends only on a very few material parameters. It also demonstrates that bolometric effect is typically much stronger than photoconductive one – the fact noticed by the experimentalist but not rigorously explained before. Based on our theory we propose and characterize new configuration of hot-electron bolometer with advantageous performance metrics.

**Optically induced change in the resistance - operation principles**

When photons (or plasmons) propagating in the waveguide get absorbed by a graphene layer their energy is transferred to the energy of photo-excited hot carriers – electrons and holes. Two processes then take place- first of all, the total density of electrons and holes increases which increases the conductivity of the graphene sheet – this is a so-called photo-conductive (PC) effect. The second effect is a bolometric effect in which the increase of the temperature causes decrease of the mobility as the electrons are moved to the higher energy states in the conduction band. We assume here an n-doped material but of course all the considerations are just as valid for a p-doped graphene.

*A. Photoconductive effect*

The rate at which energy is absorbed in graphene per unit area is (Fig. 3)

$$P_{abs}(x,y) = \pi\alpha_0 \frac{E^2(x,y,z_g)}{2\eta_0} \tag{1}$$

where $z=z_g$ defines the graphene plane, $x$ is the direction of propagation, and $\alpha_0$ is the fine structure constant. This power density is linearly related to the propagating power in the waveguide

$$P = \frac{1}{2}\iint (E \times H^*)_x dy dz \tag{2}$$

where subscript indicates projection of Poynting vector onto direction of propagation.
Multiplying and dividing eq. 1 by $P$ one obtains

$$P_{abs}(x,y) = \pi\alpha_0 P(x)/S(y) \qquad (3)$$

where the $y$-dependent effective cross-section of the waveguide is

$$\frac{1}{S(y)} = \frac{E^2(x,y,z_g)}{\eta_0 \iint (E\times H^*)_x dydz} \qquad (4)$$

The two-dimensional density of the photo-excited electron-hole pairs, shown in Fig. 1a can then be found as

$$\delta n = \delta p = P_{abs}\tau_{ee}/\hbar\omega \qquad (5)$$

where $\tau_{ee}$ is the electron-electron (e-e) scattering time which causes quick Auger-like recombination processes that lead to establishment of quasi-equilibrium. This time is very short, on the scale of tens of femtoseconds, especially in relatively highly doped graphene. The sheet conductivity of graphene is

$$\sigma = \frac{1}{\pi}\frac{e^2 E_F \tau_m}{\hbar^2} = \frac{e^2 v_F \tau_m}{\hbar}\sqrt{\frac{n}{\pi}} \qquad (6)$$

where $n$ is the doping density and $\tau_m$ is the momentum scattering time. Since both additional electrons and holes contribute to the increase in conductivity one can find the relative change in it as

$$\frac{\delta\sigma_{pc}}{\sigma} = \frac{1}{2}\frac{\delta n}{n} + \sqrt{\frac{\delta p}{n}} = \frac{1}{2}\frac{P_{abs}\tau_{ee}}{\hbar\omega n} + \sqrt{\frac{P_{abs}\tau_{ee}}{\hbar\omega n}} \qquad (7)$$

As one can see, the second term is significantly larger than the first one since the photogenerated holes are all situated near the Dirac point where the effective mobility is high. But of course the holes are not excited near the Dirac point as evident from Fig. 1a. In fact the expression eq. 6 is valid only for relatively low temperatures in comparison to Fermi energy, i.e., $E_F > k_B T$ hence the minority hole photoconductivity is not nearly as high as predicted by eq. 7 and will be estimated further on.

### B. *Bolometric effect*

Next we calculate the bolometric effect, *i.e.*, the effect of heating on conductivity in graphene. What happens is that the carriers from below Fermi level get excited to the states above Fermi level and then they quickly thermalize between themselves but not with the lattice and now they occupy states above the Fermi level. The rise of electron temperature is typically much larger than the rise of lattice temperature because the specific heat of electrons is at least two orders of magnitude less than the specific heat of lattice (phonons). Obviously there are also holes somewhat below the Fermi level as shown in Fig. 1b. The relaxation process as mentioned above is very fast and occurs on tens of femtoseconds scale due to e-e scattering.

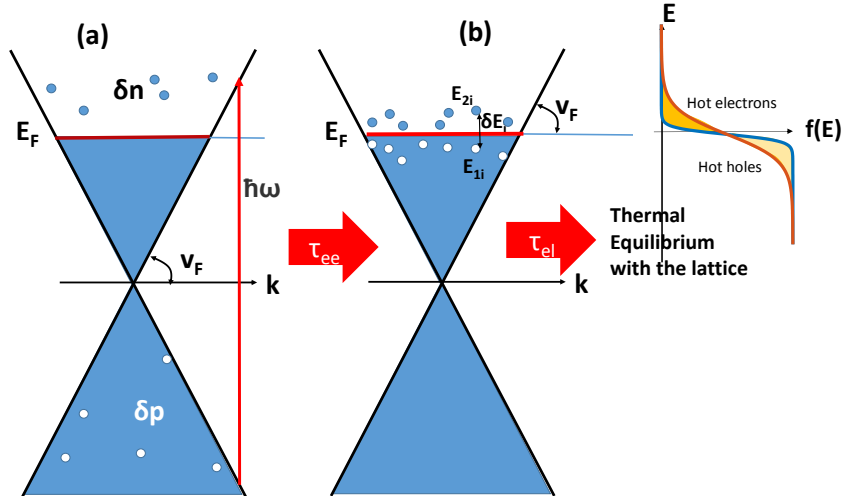

**Fig. 1.** Carrier dynamics in photo-excited graphene. (a) Immediately after photo-excitation there are extra hot electrons way above Fermi level and holes in the valence band, way below Fermi level. (b) Electron-electron scattering quickly (time $\tau_{ee}$) establishes quasi-equilibrium $f(E)$ with hot carriers in the vicinity of Femi level. The entire absorbed energy is transferred to these carriers and it stays there for a relatively long electron - lattice relaxation time $\tau_{el}$.

Even though one may introduce the effective temperature of electrons $T_e$ it is not clear that due to combination of various relaxation processes the quasi-equilibrium can be described by a perfect Fermi Dirac function. Furthermore, increased temperature will invariably cause the decrease of Fermi energy in order to preserve the total number of carriers, thus further complicating the situation. This fact should not be an obstacle to estimation of bolometric effect and here we outline how to obtain this estimate without introducing electron temperature, or in fact precise electron distribution $f(E)$. All that is necessary to know is that relative to the distribution of carriers under dark conditions, illumination causes transfer of carriers from below the original Fermi level to the states above them. In order to transfer the sate below Fermi level with energy $E_{1i}$ to the state above it with the energy $E_{2i}=E_{1i}+\delta E_i$, the energy $\delta E_i$ must be absorbed from the electro-magnetic field. Now the electrons keep energy absorbed from the field for the time $\tau_{el}$ that it takes to transfer it to the lattice phonons, and this time can be as long a 100's of femtoseconds, *i.e.*, much longer than $\tau_{ee}$. Therefore, if we perform summation over all the hot electron-hole pairs within unit area, their total energy is

$$\sum_i \delta E_i = P_{abs}\, \tau_{el} \tag{8}$$

Next we develop a model for graphene conductivity that is somewhat different from the conventional one but leads to the same result. As shown in Fig. 2 the electron in the band characterized by the wavevector $k=(k_x,k_y)$ acquires additional quasi-momentum in the presence of DC electric field $F$ applied along x direction.

$$\delta k = -eF\tau_m/h \tag{9}$$

where $F$ is the electric field and $\tau_m$ is previously-introduced momentum scattering time.

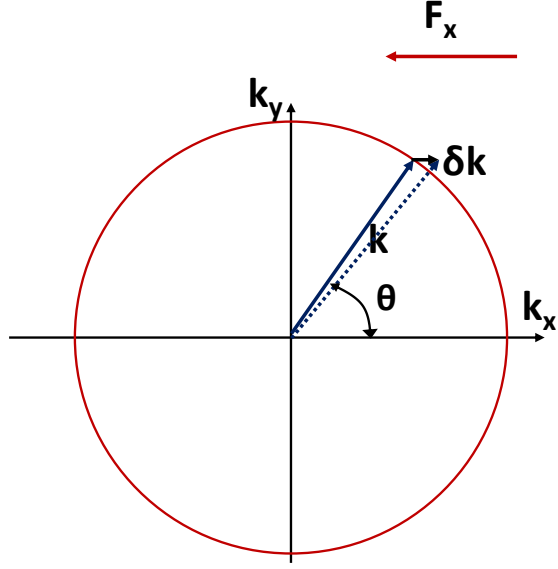

**Fig. 2.** Electron dynamics in the presence of optical field *F*.

The projection of the velocity of this electron onto the direction of field is

$$v_x = v_F \cos\theta = v_F \frac{k_x}{\sqrt{k_x^2 + k_y^2}} \tag{10}$$

Taking derivative, we obtain

$$\frac{dv_x}{dk_x} = \frac{v_F k_y^2}{k^3} = \frac{v_F}{k}\sin^2\theta \tag{11}$$

Therefore, the velocity of the electron in the presence of DC field is

$$v_x = v\cos\theta - \frac{ev_F \tau_m}{hk} F \sin^2\theta \tag{12}$$

Now one can obtain the expression for the current density as

$$J_x = -e \sum_{|k|\le k_F} ev_x = \frac{g_v}{2\pi^2} \frac{e^2 v_F \tau_m}{h} F \int_0^{2\pi}\int_0^{k_F} k^{-1}\sin^2\theta \, d\theta \, dk = \frac{1}{\pi}\frac{e^2 E_F \tau_m}{h^2} F \tag{13}$$

where $g_v$=2 is the valley degeneracy, which yields precisely eq. 6 for the sheet conductivity. Now, if we average eq. 12 over the angle $\theta$ we can write

$$\langle v_x \rangle = -\frac{e v_F^2 \tau_m}{2E} F = -\mu F \tag{14}$$

where the energy-dependent mobility is

$$\mu(E) = \frac{e v_F^2 \tau_m}{2E} \tag{15}$$

and

$$\sigma = e \sum_k f(k)\mu(E) \quad (16)$$

where *f(k)* is the distribution function. Herein lies the physical origin of the strength of bolometric effect in graphene – the extreme non-parabolicity of the dispersion. If the dispersion had been parabolic, then the square of velocity in eq. 15 would have been proportional to energy as $m_c v^2/2 = E$, where $m_c$ is effective mass, which would have made mobility independent of energy, $\mu = e\tau_m/m_c$.

Therefore, when the *i-th* electron gets promoted from state 1*i* to state 2*i* (leaving a hole in state 1*i* behind) due to heating, its mobility gets reduced by

$$\delta\mu_i = \frac{\partial \mu(E)}{\partial E}\bigg|_{E_F} \times \delta E_i = -\mu(E_F)\frac{\delta E_m}{E_F} = -\frac{ev_F^2 \tau_m}{2E_F^2}\delta E_i \quad (17)$$

All that is left is to calculate the total bolometric change of sheet conductivity by summing up individual electron-hole pair contributions eq. 17 within unit area according to eq. 16 and use eq. 8

$$\delta\sigma_b = e\sum_i \delta\mu_i = -\frac{e^2 v_F^2 \tau_m}{2E_F^2}\sum_i \delta E_i = -\frac{e^2 v_F^2 \tau_m}{2E_F^2} P_{abs}\tau_{el} \quad (18)$$

From eq. 6 we can express $e^2\tau_m = \sigma\pi\hbar^2/E_F$ and therefore relative bolometric change is

$$\frac{\delta\sigma_b(x,y)}{\sigma} = -\frac{\pi\hbar^2 v_F^2}{2E_F^3}P_{abs}(x,y)\tau_{el} = -\frac{P_{abs}(x,y)\tau_{el}}{2E_F n} \quad (19)$$

The change in resistance can then be simply estimated by averaging eq. 19 as

$$\frac{\Delta R_b}{R} = \frac{\int_{-L_1/2}^{L_1/2}\int_0^W P_{abs}(x,y)dxdy}{2E_F nWL_1}\tau_{el} = \frac{\eta_{abs}P\tau_{el}}{2E_F nWL_1} \quad (20)$$

where *W* and *L₁* are the width and the length of graphene that absorb a light and $\eta_{abs}$ is the absorption efficiency. Note that if the doping is electrostatic then the total charge in the graphene can be found as roughly $enWL = C_g(V_g - V_t)$ where $C_g$ is gate capacitance and $V_t$ is threshold voltage and then

$$\frac{\Delta R_b}{R} = \frac{\eta_{abs}eP\tau_{el}}{2E_F C_g(V_g - V_t)} \quad (21)$$

C. *Comparison of two effects*

Before continuing we need to develop the expression for the hole conductivity for the case of low carrier concentration, when the Fermi energy is comparable or less than $k_B T$ or $\delta p < N_{eff} = (2/\pi)(k_B T/\hbar v_F)^2 \sim 10^{10} cm^{-2}$ and the carriers are distributed according to Boltzmann distribution with energy counted down

$$f(E) = exp\frac{E_F - E_V}{k_B T_p} \quad (22)$$

where $E_V$ is the quasi-Fermi level of valence band and $T_p$ is temperature of hot holes following excitation, which is, in general photon energy dependent and can be anywhere from lattice temperature to $\hbar\omega/2k_B$.

We only need a very rough idea about its magnitude in order to show that PC effect is weaker than bolometric one. Substituting eq. 22 into eq. 16, we then obtain the expression for the holes photo-conductivity

$$\delta\sigma_p = e \frac{\delta p}{k_B^2 T_p^2} \int \mu(E) e^{-\frac{E}{k_B T_e}} E dE = \frac{e^2 v_F^2 \tau_m}{2 k_B T_p} \delta p \tag{23}$$

Using eq. 6 one obtains

$$\frac{\delta\sigma_p}{\sigma} = \frac{\pi h^2 v_F^2}{2 k_B T_p E_F} \delta p = \frac{E_F}{2 k_B T_p} \frac{\delta p}{n} \tag{24}$$

Since $\delta n = \delta p$ we obtain

$$\frac{\delta\sigma_{pc}}{\sigma} = \frac{1}{2} \frac{\delta n}{n} \left(1 + \frac{E_F}{k_B T_p}\right) = \frac{1}{2} \frac{P_{abs} \tau_{ee}}{\hbar \omega n} \left(1 + \frac{E_F}{k_B T_p}\right) \tag{25}$$

We can now compare photoconductive response with the bolometric one – essentially

$$-\frac{\delta\sigma_{pc}}{\delta\sigma_b} = \frac{E_F \tau_{ee}}{\hbar \omega \tau_{el}} \left(1 + \frac{E_F}{k_B T_p}\right) \ll 1 \tag{26}$$

This follows from the fact that $E_F < \hbar\omega/2$, $\tau_{ee} < \tau_{EL}$, and the expression in parenthesis is on the order of a few times unity. Hence bolometric effect dominates.

**Bolometric photodetector arrangement**

To maximize the PB photodetector performances expressed by eq. 20 and 21, we propose a plasmonic bolometric photodetector operating at telecom wavelengths that is based on the long-range dielectric loaded surface plasmon polariton (LR-DLSPP) waveguide [48-50] with graphene placed at the maximum electric field of the propagating mode (Fig. 3). As a result, the plasmonic mode is highly absorbed by the graphene sheet enhancing thermally activated carrier transport in the graphene [44] while the absorption losses in metal are highly reduced.

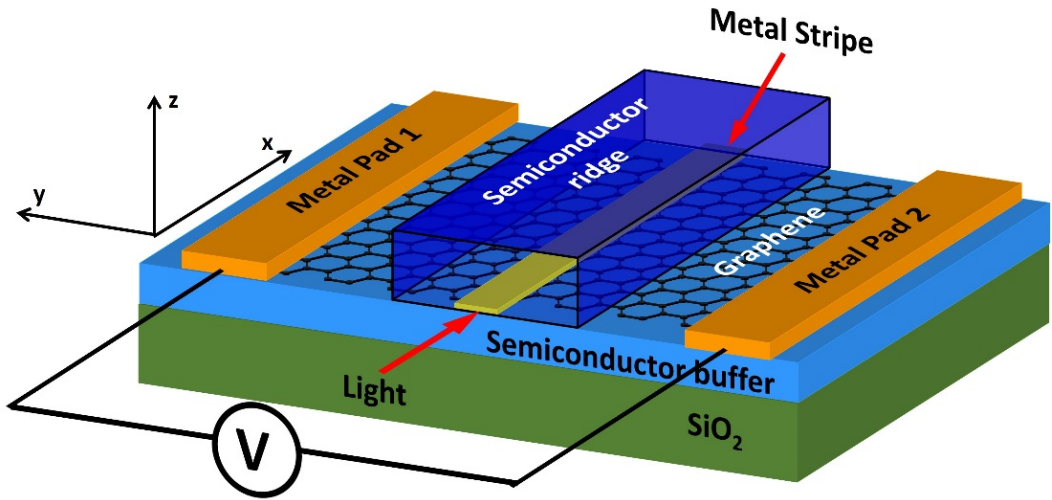

**Fig. 3.** Schematic of the proposed plasmonic PB photodetector in symmetric external electrodes arrangement. Here, Si is used as both Semiconductor ridge and Semiconductor buffer material, however any other material can be implemented in this design. Here, a distance between Metal Pad 1 and Metal Pad 2 defines the length of the photodetector *L* (along *y* axis direction), while the length of the waveguide defines the width of the photodetector *W* (along *x* axis direction).

To achieve efficient absorption in graphene it is necessary to maximize the in-pane component of electric field. The in-plane electric field component of the propagating LR-DLSPP mode interacts strongly with graphene enhancing absorption [Fig. 4]. The presence of the in-plane component of the electric field even for the TM mode is associated here with the small thickness of the metal stripe and its sharp metal corners [31]. The electric field at the metal stripe's corners is very strong but decays quickly on the graphene. This produces hot carriers in very close proximity to the metal stripe.

To minimize the contribution from the photo-thermoelectric (PTE) effect, that requires an asymmetric electron distribution in a graphene channel [13, 16, 33], the symmetric contact arrangement has been implemented here, *i.e.*, the same metals were used as contacts and the structure was symmetric with respect to the center of the metal stripe/ridge [Fig. 3]. As a result, a symmetric band diagram across the active graphene channel is achieved. Consequently, only the aforementioned photo-bolometric (PB) and photo-conductive (PC) effects can exist under this arrangement and, as has been shown above, PC effect is invariably weaker than a bolometric one.

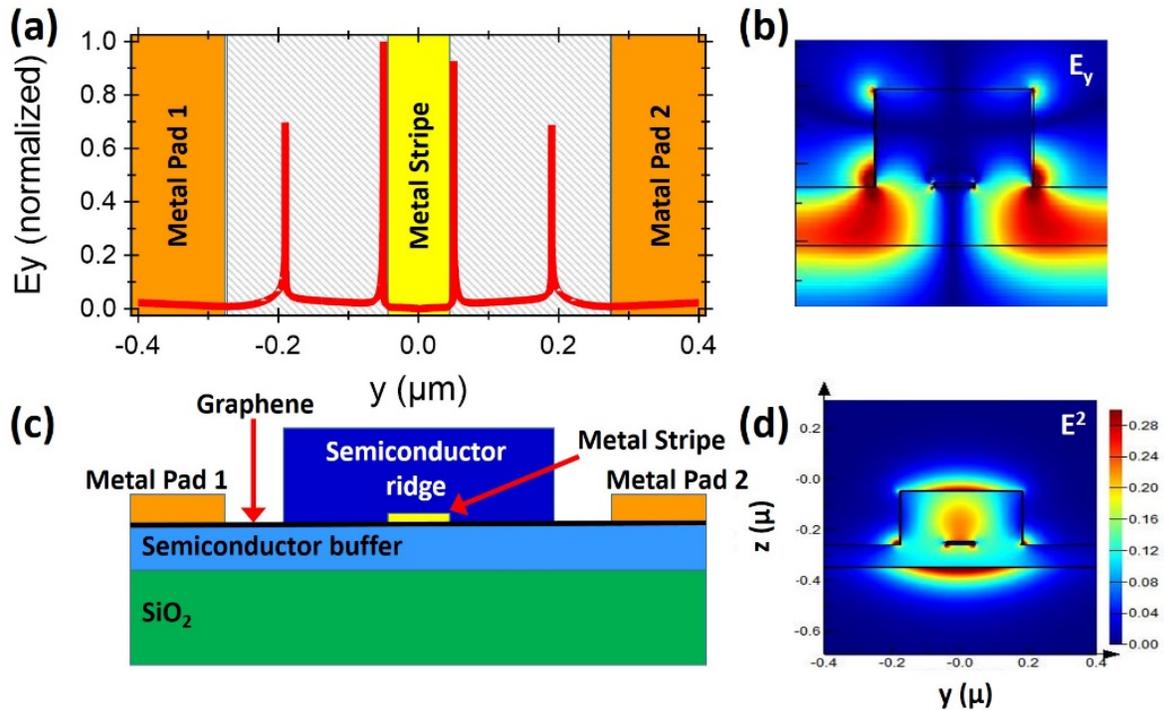

**Fig. 4.** (a) In-plane electric field component of the plasmonic TM mode on graphene surface with a corresponding (b) electric field profile ($E_y$) for the cross-section of the photodetector as presented in (c). The $E^2$ of the mode is presented in (d).

**Electron-electron and electron-phonon scattering times**

The theory of hot-carrier induced bolometric effect presented in previous sections relies on the fact that carriers inside the graphene stay out of equilibrium with lattice for a relatively long time $\tau_{el}$ [51-53] that is longer than the time that it takes to establish the thermalized distribution among the electrons - $\tau_{ee}$

[54-56]. Here we preset estimates of these rates and compare them with available experimental data. The e-e scattering rate, under large doping of graphene, when the Fermi energy $E_F$ exceeds $k_B T_e$, the inelastic scattering rate tends to form

$$\tau_{ee}^{-1} \sim \alpha^2 \frac{(k_B T_e)^2}{\hbar E_F} \tag{27}$$

Here, $T_e$ is the electron temperature in graphene and $\alpha$ is defined as [57]

$$\alpha = \frac{e^2}{\epsilon_r \hbar v_F} \sim 1 \tag{28}$$

where $\epsilon_r$ is the dielectric constant. In the next step, the energy from the electrons is dissipated (relaxed) either by the energy transfer to the lattice via the e-l coupling or by electron diffusion away from the heated region [33, 47, 58] (Fig. 5). The e-l scattering rate is defined as [59]

$$\tau_{el}^{-1}(E_k) = \frac{(E_k - E_F)D^2}{4\rho_m v_S^2 \hbar^3 v_F^2} k_B T_L \tag{29}$$

where $D$=18 eV is the deformation potential, $\rho_m$=7.4·10$^{-7}$ kg/m$^2$ is the mass density of monolayer graphene [39, 60], $v_s$=2.6·10$^4$ m/s is the phonon velocity in graphene, $E_k$ is the carrier energy and $T_L$ is the lattice temperature.

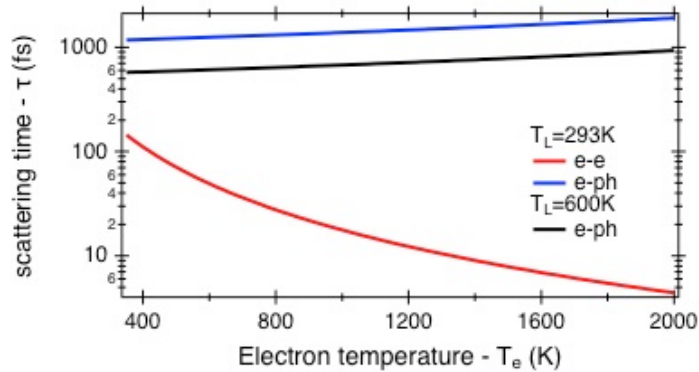

**Fig. 5.** Electron temperature dependent e-e and e-l scattering times for different lattice temperature $T_L$=293 K (red and blue curves) and $T_L$=600 K (black curve). The ambient (room) temperature was assumed at $T_0$=293 K.

In graphene, the Fermi energy is much smaller than in conventional metals, thus the e-e scattering time is extremely fast even is the doped regime with the smallest value achieved near the Dirac point [57]. In comparison, the e-l scattering time is much slower than e-e scattering time and range from 1 ps to 2 ps under the room temperature operation conditions [33, 36, 47, 51, 54, 55]. However, for higher lattice temperature $T_L$, the e-l scattering time decreases (51, 56) (Fig. 5). From the scattering rates it can be concluded that an increase of phonon/lattice temperature $T_L$ leads to increased e-l scattering rate, *i.e.*, decreased e-; scattering time and, in consequence, a reduction in transport current, *i.e.*, decrease of $\Delta R/R$, while an increase in electron temperature $T_e$ is equivalent to an increased carrier density and leads to an increase in transport current. The increase of the lattice temperature $T_L$ leads to more efficient cooling pathway for hot electrons [53], since additional phonons become available for heat dissipation [25, 26, 54, 61]. The lattice temperature can be increased in the graphene either through a

Joule heating that is proportional to the applied electrical power [25, 54, 61] or through highly confined plasmonic energy (26). Thus, for compact devices, where the (electrical) power density is high, the increases of $T_L$ can provide additional heat dissipation channel, in consequence, reducing a PB effect [32]. Simultaneously, as it has been previously observed [54], the efficiency of electron heating is independent of lattice temperature and depends only on the in-plane component of the electric field coupled to the graphene.

**Evaluation of PB photodetector performances**

Having estimated the scattering times, we can evaluate the detector performance. For a large distance of hot electrons to external electrodes exceeding the electron mean free path $l_{MFP}$, the main heat-flow channel of hot electrons is through relaxation to the graphene lattice. In such a case, the ratio of resistance $\Delta R/R$ is very valuable formula to evaluate a bolometric photodetector. The ratio of resistance $\Delta R/R$ is calculated from eq. 20 and results are showed in Fig. 6 with $W$=40 μm and $L_1$=10 nm. As observed from Fig. 6, the maximum resistance ratio was calculated at $\Delta R/R$=23 for $E_F$=0.1 eV. When compared with other plasmonic PB photodetectors [12], the resistance ratio was much smaller calculated at $\Delta R/R$=7.7 for $E_F$=0.1 eV.

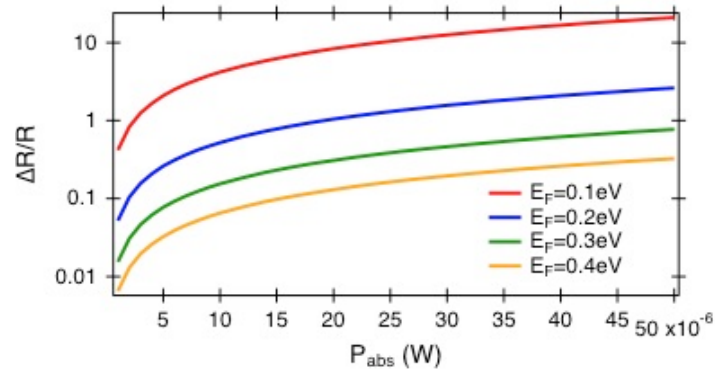

**Fig. 6.** Ratio of resistances as a function of absorbed power by graphene sheet for different Fermi energies $E_F$, carrier concentrations $n$. Here, the lattice temperature was kept at the ambient temperature, *i.e.*, $T_L$=$T_0$=293 K.

To evaluate $\Delta R/R$, the power absorbed by the graphene should be calculated. The power absorbed by the graphene photodetector $P_{abs}$ is related to the input power $P_{in}$ as $P_{abs}=\eta_{abs}\eta_c P_{in}$, where $\eta_c$ and $\eta_{abs}$ are the coupling and absorption efficiencies, respectively. Here $L_1$ is the length of the in-plane electric field interacting with graphene. The coupling efficiency in this type of plasmonic waveguide can exceed 90 %, with absorption efficiency exceeding 40 % for 40 μm-long and up to 63 % for 100 μm-long photodetectors, respectively (Fig. 7) (Methods section). Simultaneously, as showed in Fig. 4a, the power is absorbed by less than 10 nm-wide graphene. As a result, the power absorbed by the graphene is extremely high, enhancing the electron temperature in the graphene.

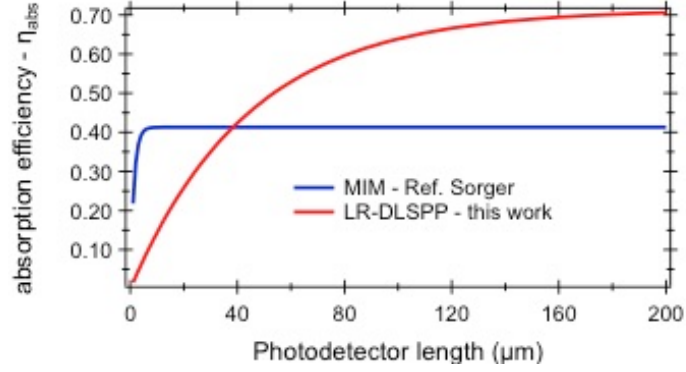

**Fig. 7.** (a) Comparison of the light absorption efficiencies in graphene for proposed here LR-DLSPP-based photodetector and MIM-based photodetector [32].

Under zero bias voltage, the PB photocurrent does not exist. However, under a bias voltage applied across the metallic contact, the change of graphene resistance can be detected by the change of the photocurrent flowing through the graphene sheet as

$$I_{ph} = \Delta I = \frac{V_{bias}}{\Delta R} \tag{32}$$

Thus, the responsivity of photodetector is expressed by

$$R_{ph} = \frac{I_{ph}}{P_{in}} \tag{33}$$

Consequently, knowing the photocurrent of the photodetector and input power, the responsivity can be calculated. The calculations were performed for conductivity $\sigma_0$=0.4 mS, similar to Ref. 12. For $L$=800 nm long and $W$=40 µm wide photodetector the resistance was calculated at $R$=50 Ω. Thus, by the applying a bias voltage of 1 V, a current $I$=20 mA at room temperature $T_0$=293 K was calculated. As observed from Fig. 6 (ratio of resistances), the PB photodetector works in inverse operation mode with the off-state in the dark (where the current signal is high) and the on-state with light incidence (where the current signal is low). Furthermore, to achieve a large on-off state a strong suppression of the current is highly desired with an applied optical signal. This observation is consistent with experimental work performed with the bow-tie photo-bolometric photodetector [12]. The current change between off and on state corresponds to a photocurrent (eq. 32).

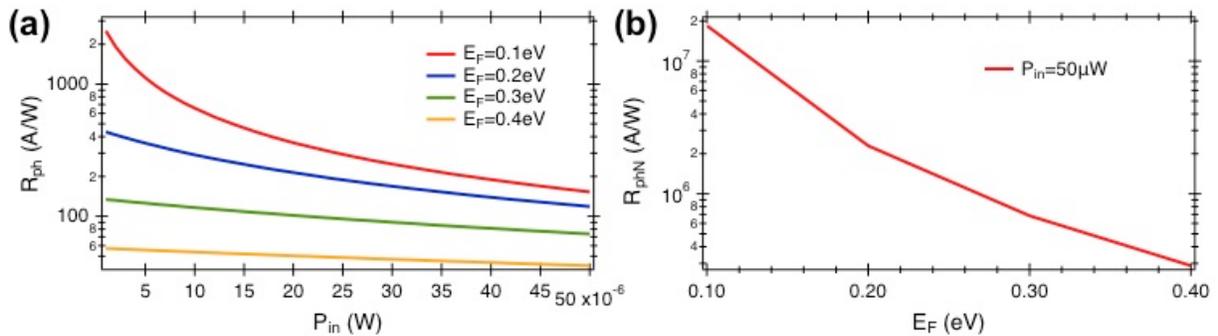

**Fig. 8.** (a) Responsivity as a function of input power $P_{in}$ for different Fermi energies and (b) normalized current responsivity as a function of Fermi energy $E_F$.

When light is delivered to the photodetector with input power of 50 µW, that corresponds to the absorbed power of 20 µW, the external responsivity was calculated at $R_{ph}$=150 A/W, and $R_{ph}$=120 A/W for Fermi energies of 0.1 eV, and 0.2 eV (Fig. 8a). For lower input power of 5 µW, *i.e.*, absorbed power $P_{abs}$=2 µW, the external responsivity was calculated at $R_{ph}$=1100 A/W for $E_F$=0.1 eV, while for $E_F$=0.2 eV it was calculated at $R_{ph}$=350 A/W (Fig. 8a). As observed from above, the low power operation is desired for the best performance of the photo-bolometric photodetector as it reduces power requirements and ensures enhances responsivity for all Fermi energies studied here.

Another very valuable parameter that characterize the performance of the photodetectors is the normalized current responsivity (in %/W) defined as [62]

$$R_{ph,N} = \frac{\frac{I_{ON} - I_{OFF}}{I_{OFF}} \cdot 100\%}{P_{in}} \tag{34}$$

where $I_{ON}$ is the current without illumination and $P_{in}$ is the input power. For $\Delta R/R \approx 10$ at $E_F$=0.1 eV and for $P_{abs}$=20 µW absorbed power (Fig. 6), that corresponds to the input power of 50 µW, the normalized current responsivity was calculated at $R_{ph,N}$=1.9·10$^7$ %/W (Fig. 8b). It is three orders of magnitude higher than state-of-the art pyroelectric bolometer based on graphene-lithium niobate (LN), $R_{ph,N}$=2·10$^4$ %/W [62], and five orders of magnitude higher than pyroelectric bolometer based on a graphene-lead zirconate titanate (PZT), $R_{ph,N}$=1.2·10$^2$ %/W [63]. However, despite of a comparable responsivity of the graphene-lithium niobate photodetector, the operation speed is limited to 1 kHz what makes it inefficient for high speed operations.

In graphene hot electron photodetectors such as PB photodetectors, the response time and thus the bandwidth of the photodetector are determined by the e-l relaxation time that is required for graphene devices to return to equilibrium [10, 25, 26, 33, 53]. As previously experimentally observed [25, 26, 33, 51, 53, 56, 64] and confirmed by calculations performed here, the e-l relaxation time range from hundreds of femtoseconds to tens of picoseconds that depends on the carrier concentration in graphene as well as the lattice and electron temperatures. As a consequence, graphene hot electron bolometers enable a realization of photodetectors with the bandwidth approaching 1 THz.

**Conclusion**

Here, a theory of bolometric effect originating from the band nonparabolicity of graphene was developed and new waveguide-integrated plasmonic graphene bolometer was proposed with the ratio of resistances exceeding 23. As such, the responsivity exceeding 1100 A/W is expected to be attained .The extremely fast response time of hot carriers in graphene enable a realization of photodetectors with response well beyond hundreds of GHz. The improved performances originate from the highly localized in-plane component of the electric field that is mostly absorbed within 10 nm from a metal stripe. The results show the potential of graphene for high-speed communication systems.

**Methods**

**Estimation of power absorbed by graphene sheet**

To evaluate the performance of the proposed photodetector the amount of power absorbed by a graphene sheet need to be determined. Power absorbed by the graphene sheet can be calculated as follow

$$P_{abs} = P_{in} e^{-\alpha \cdot L} \tag{35}$$

Here we simulate around $P_{in}$=50 µW of power coupling to a photodetector that consist of Si as Semiconductor ridge and Semiconductor buffer and Au as Metal stripe (Fig. 3 and 4). The Si ridge (Semiconductor ridge) width and thickness were takes at 380 nm and 210 nm, respectively while Si rib thickness (Semiconductor buffer) was taken at 90 nm. The Au stripe width was taken at 80 nm and thickness at 12 nm, while a distance between external electrodes (Metal Pad 1 and Metal Pad 2) was taken at 800 nm.

For the absorption coefficient of graphene $\alpha_G$ and metal $\alpha_M$ obtained from a simulation, the length-dependent of fraction of light absorption in graphene $\eta_{abs}$ can be calculated by:

$$\eta_{abs} = \frac{\alpha_G}{\alpha_M + \alpha_G}(1 - e^{-\alpha_G L} e^{-\alpha_M L}) \qquad (36)$$

where $L$ is the length of photodetector. Calculations were performed for a telecom wavelength of 1550 nm and results were summarized in Fig. 7. From this figure, it can be deduced that for a 40 µm-long photodetector about 40 % of the power is absorbed by graphene (Fig. 7). As a result, we can assume a power of $P_{abs}$=20 µW absorbed by graphene that will contribute to a photocurrent generation.

## Author information

**Affiliations**

New York University Abu Dhabi, Saadiyat Island, PO Box 129188, Abu Dhabi, UAE

Jacek Gosciniak

John Hopkins University, Baltimore, MD 21218, USA

Jacob Khurgin

**Contributions**

J.G. conceived the idea of detector, performed all calculations, FEM, FDTD simulations and wrote the manuscript. J.B.K derived the analytical estimate of bolometric and PC responses. J.G. and J.B.K analyzed and discussed the results. Both authors reviewed the manuscript.

**Corresponding author**

Correspondence to Jacek Gosciniak (jeckug10@yahoo.com.sg)